\documentclass[11pt,twoside]{article}
\usepackage{graphicx,latexsym,fullpage,jeffe,url}
\newcommand{\pict}[2]{\centerline{\includegraphics[scale=#1]{#2.eps}}}

\newtheorem{theorem}{Theorem}

\newtheorem{conjecture}[theorem]{Conjecture}

\begin{document}

\def\threesum{\textsc{3sum}}
\def\fakethreesum{3{\scriptsize SUM}}  
\def\threesump{\threesum$'$}
\def\threesumprep{\threesum\textsc{Prep}}

\urldef{\Jurl}\url{http://www.cs.uiuc.edu/~jeffe/}
\urldef{\Murl}\url{http://www.cs.uu.nl/people/markov/}


\title{Preprocessing Chains for Fast Dihedral Rotations\\
	 Is Hard or Even Impossible}

\author{
	Michael Soss\thanks{
		School of Computer Science, McGill University.
		Present Address: Chemical Computing Group, Montreal;
		soss@chemcomp.com}
	\and 
	Jeff Erickson\thanks{
		Department of Computer Science, University of Illinois
		at Urbana-Champaign; jeffe@cs.uiuc.edu;	\Jurl .
		Research supported in part by a Sloan Fellowship.}
	\and 
	Mark Overmars\thanks{
		Institute of Information and Computing Sciences,
		Utrecht University; markov@cs.uu.nl; \Murl .}
} 

\date{Submitted to \emph{Computational Geometry:~Theory and
	Applications}:~ \today}
\maketitle

\begin{abstract}
We examine a computational geometric problem concerning the structure
of polymers.  We model a polymer as a polygonal chain in three
dimensions.  Each edge splits the polymer into two subchains, and a
\emph{dihedral rotation} rotates one of these subchains rigidly about
the edge.  The problem is to determine, given a chain, an edge, and an
angle of rotation, if the motion can be performed without causing the
chain to self-intersect.  An $\Omega(n\log n)$ lower bound on the time
complexity of this problem is known.

We prove that preprocessing a chain of $n$ edges and answering $n$
dihedral rotation queries is \threesum-hard, giving strong evidence
that $\Omega(n^2)$ preprocessing is required to achieve sublinear
query time in the worst case.  For dynamic queries, which also modify
the chain if the requested dihedral rotation is feasible, we show
that answering $n$ queries is by itself \threesum-hard, suggesting
that sublinear query time is impossible after \emph{any} amount of
preprocessing.
\end{abstract}


\section{Introduction}

During the past several decades, questions regarding polymer structure
have received widespread interest in the physics community.
Throughout the literature, a polymer is often modeled as a
self-avoiding chain of line segments in three-space, where the
vertices represent atoms and the edges represent bonds.  Due to the
constraints of a chemical bond, the valence angles---angles between
adjacent bonds to the same atom---are often fixed to attain a more
realistic model~\cite{benoit-48, eyring-32, kuhn-34, taylor-48},
resulting in a limited range of motion.

\begin{figure}[t]
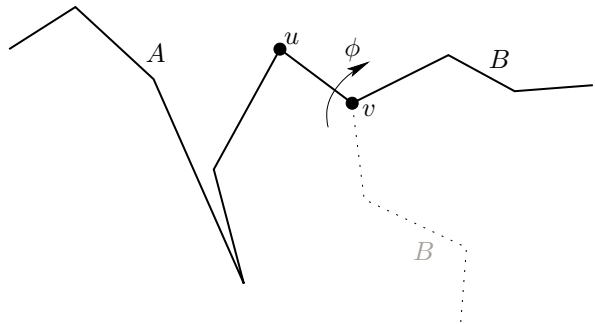

\centerline{\input edgespin-ltx.pstex_t}
\caption{A dihedral rotation.}
\label{fig-edgespin}
\end{figure}

\begin{figure}[t]
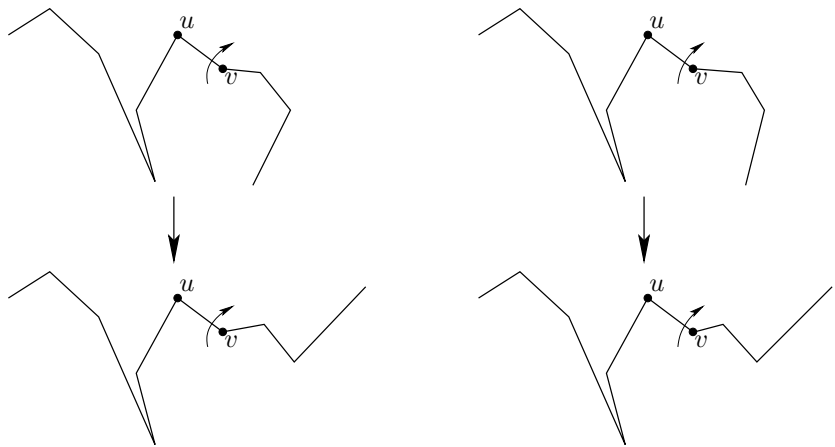

\centerline{\input stereo-flatedgespin-ltx.pstex_t}
\caption{A dihedral rotation, shown as a stereogram.  The image can be
viewed in stereo by crossing one's eyes until the arrows coincide.}
\label{fig-stereo-edgespin}
\end{figure}

The most common method to sample the configuration space of polymers
is to randomly reconfigure the chain in a Monte Carlo
simulation~\cite{curro-74, curro-76, freire-76, mckenzie-76,
stellman-72b, stellman-72}.  An edge of the chain is chosen at random,
and a \emph{dihedral rotation} is performed.  Any edge $\overline{uv}$
splits the chain into two subchains $A$ and $B$, where $u \in A$ and
$v \in B$.  A dihedral rotation at $\overline{uv}$ rotates the
subchain~$B$ rigidly by some angle $\phi$ (or equivalently, rotates
$A$ by angle $-\phi$) around $\overline{uv}$, keeping the angles at
$u$ and $v$ fixed.  See Figures~\ref{fig-edgespin}
and~\ref{fig-stereo-edgespin}.

Before each dihedral rotation, the simulation must check whether the
motion is \emph{feasible}, that is, whether or not the chain
intersects itself at any time during the motion.  Because
self-intersections are not allowed in the model, if a rotation is
deemed infeasible the resulting configuration must be discarded and
another motion randomly chosen.  The probability that a randomly
selected motion is feasible decreases rapidly as larger polymers are
considered.  Thus, it is important to determining whether or not a
dihedral rotation is feasible as quickly as possible.  Soss and
Toussaint~\cite{st-gcapr-00} formalized this problem as follows.

\begin{problem}{Dihedral Rotation}
Given a polygonal chain, an edge $\overline{uv}$ of the chain, and an
angle $\phi$, is the dihedral rotation of angle $\phi$ at
$\overline{uv}$ feasible?
\end{problem}

Soss and Toussaint proved an $\Omega(n \log n)$ bound on the time
complexity of this problem and described a brute force algorithm that
runs in $O(n^2)$ time and $O(n)$ space, where $n$ is the number of
edges in the chain.  For the special case where $\phi \geq 2\pi$ (a
full rotation), they constructed a faster algorithm with the help of
results by Agarwal and Sharir~\cite{as-rbida-90} and by Guibas,
Sharir, and Sifrony~\cite{gss-gmppt-89} on arrangements of curves.
This algorithm runs in expected time $O(n 2^{\alpha(n)} \log n)$,
where $\alpha(n)$ is the slowly-growing inverse Ackermann function.


These results apply to single motions, but the simulation of a polymer
is a complex process.  A typical simulation might have hundreds or
thousands of attempted motions.  In this paper we examine the
complexity of computing the feasibility of a sequence of dihedral
rotations.  We will refer to each such determination as a
\emph{dihedral rotation query}.  We will distinguish between
\emph{static} queries, which do not modify the chain, and
\emph{dynamic} queries, which actually perform the dihedral rotation
if it is feasible.  To compute each motion as if it were a separate
problem seems inefficient as the chain always maintains its edge
lengths and vertex-angles.  Thus, an intuitive goal is to preprocess
the chain so that each ensuing dihedral rotation query can be solved
in $o(n \log n)$ time.

We show two problems concerning multiple dihedral rotations to be
\emph{\fakethreesum-hard}.  A problem is \threesum-hard if there is a
subquadratic reduction from the following problem.

\begin{problem}{\fakethreesum}
Given a set of integers, do any three elements sum to zero?
\end{problem}

\threesum-hardness was introduced by Gajentaan and
Overmars~\cite{go-copcg-95-fixed} to provide evidence in support of
conjectured $\Omega(n^2)$ lower bounds for several problems.  The best
known algorithm for \threesum\ runs in time $\Theta(n^2)$.  Quadratic
lower bounds have been proven for \threesum\ and a few other
\threesum-hard problems in restricted models of
computation~\cite{e-lblsp-98,e-nlbch-99,es-blbda-95}, but the
strongest lower bound for any of these problems in a general model of
computation is $\Omega(n\log n)$, which follows from results of
Ben-Or~\cite{b-lbact-83}.

In Section~\ref{sec-sesq}, we consider \emph{static} dihedral rotation
queries, which determine whether a given dihedral rotation is feasible
or not, without modifying the chain.  We show that preprocessing the
chain and answering $n$ static dihedral rotation queries is
\threesum-hard.  Thus, $\Omega(n^2)$ preprocessing is almost certainly
required to achieve sublinear query time.

In Section~\ref{sec-mesq}, we consider \emph{dynamic} dihedral
rotation queries, which either modify the chain by performing a
dihedral rotation or report that the desired rotation is infeasible.
We show that dynamic dihedral rotation queries cannot be answered in
sublinear time after \emph{any} amount of preprocessing, unless there
is a \emph{nonuniform} family of algorithms for \threesum\ with
subquadratic running time.  Since this seems unlikely, especially in
light of existing lower bounds~\cite{e-lblsp-98}, answering a single
dynamic dihedral rotation query almost certainly requires $\Omega(n)$
time in the worst case.  Even if such a nonuniform family of
algorithms does exist, the preprocessing time would be at least the
time required to construct the $n$th algorithm in the family.  In
contrast, if we do not need to check for feasibility, we can perform
any dihedral rotation in $O(\log n)$ time, after only $O(n)$
preprocessing.


\section{Static dihedral rotation queries}
\label{sec-sesq}

In this section, we consider the problem of preprocessing a chain of
$n$ segments so that we can quickly determine whether an arbitrary
dihedral rotation is feasible.  We refer to such tests as
\emph{static} dihedral rotation queries because they only test
feasibility; performing a query does not actually modify the chain.
We consider \emph{dynamic} queries, which either modify the chain or
report a collision, in the next section.

We are interested in tradeoffs between the preprocessing time and the
worst-case query time.  For example, using the algorithm of Soss and
Toussaint~\cite{st-gcapr-00}, we can compute the degrees of freedom
for every possible dihedral rotations in $O(n^3)$ time; if we store
the results in a table, then any query can be answered in $O(1)$ time
simply by looking up the result.  On the other hand, with no
preprocessing, the optimal query time lies somewhere between
$\Omega(n\log n)$ and $O(n^2)$~\cite{st-gcapr-00}.

The remainder of this section provides strong evidence for the
following conjecture.

\begin{conjecture}
\label{conj-lbedgespinquery}
In any scheme to preprocess a chain of $n$ edges to answer static
dihedral rotation queries, either the preprocessing time is
$\Omega(n^2)$ or the worst-case query time is $\Omega(n)$.
\end{conjecture}

We provide strong support for this conjecture by proving that
preprocessing a chain of $n$ segments and performing $n$ static
dihedral rotation queries is \threesum-hard.  To simplify our
reduction, rather than using \threesum\ directly, we will instead use
the following closely related problem.

\begin{problem}{\fakethreesum$'$}
Given three sets of integers, are there elements, one from each set,
whose sum is zero?
\end{problem}

Using \threesump\ instead of \threesum\ poses no additional
complication, since the two problems are reducible to one another in
linear time, only changing the complexity of the input by a constant
factor~\cite{go-copcg-95-fixed}.  Therefore a reduction from
\threesum\ is equivalent to a reduction from \threesump.

Because the time complexity of \threesump\ is unknown, we will use the
notation $\threesum(n)$ to denote the time complexity of the
\threesump\ problem, where $n$ is the total size of the three sets.

\begin{theorem}
\label{theo-lbedgespinquery}
Preprocessing a chain of $n$ edges and performing $n$ static dihedral
rotation queries is \threesum-hard.
\end{theorem}

\begin{proof}
Given any instance of \threesump, we create a polygonal chain of $n$
segments in $O(n\log n)$ time, such that a sequence of $O(n)$ dihedral
rotation queries solves the \threesump\ problem.  Thus, if we spend
$P(n)$ time preprocessing the chain and $Q(n)$ time answering each
query, then $P(n) + nQ(n) = \Omega(\threesum(n))$.

We begin by modifying the sets so that each set lies in an interval
far from the other sets.  Specifically, we replace $A$ and $C$ with
two new sets $A' = \set{a - 2m \mid a\in A}$ and $C' = \set{c + 2m
\mid c\in A}$, where $m$ is the maximum absolute value of any element
in $A \cup B \cup C$.  This replacement clearly does not affect the
outcome of \threesump.  To simplify the reduction, we also sort the
three sets in $O(n\log n)$ time.  (There is a more complicated
$O(n)$-time reduction that avoids sorting by exploiting the third
dimension.)

We create a planar chain as illustrated in Figure~\ref{fig-threesum}.
The chain consists of two \emph{combs} joined by an axis-parallel
\emph{staircase}.  For each element $a'\in A'$, the left comb contains
a very slim upward tooth centered on the line $x=a'$.  For each
element $c'\in C'$, the right comb contains a very slim downward tooth
centered on the line $x=c'$.  Finally, for each element $b\in B$, the
staircase contains a vertical edge on the line $x=-b/2$.

\begin{figure}
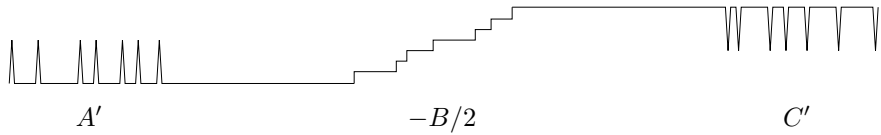

\centerline{\input threesum.pstex_t}
\caption{Reducing \threesump\ to a series of static dihedral rotation
queries.}
\label{fig-threesum}
\end{figure}

We now ask a series of $O(n)$ static dihedral rotation queries;
namely, can a dihedral rotation of angle $2\pi$ be performed at each
vertical edge in the orthogonal staircase?  Since the edge is
vertical, and the chain is planar, the only possibility for an
intersection is when the rotation has reached $\pi$.  At this point,
one comb and part of the staircase have been reflected across the
vertical edge, as in Figure~\ref{fig-examplethreesumspin}.

\begin{figure}
\pict{0.35}{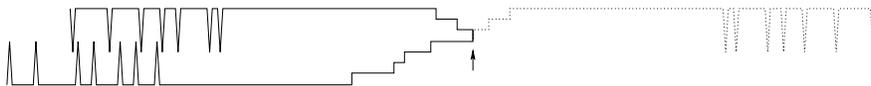}
\caption{A dihedral rotation at a vertical staircase edge.}
\label{fig-examplethreesumspin}
\end{figure}

Because the rotation is performed at a vertical edge, no edge changes
height.  This immediately implies that the staircase cannot
self-intersect.  Each comb stays individually rigid, so neither comb
can self-intersect.  Furthermore, because each vertical edge in the
staircase is at distance at most $m$ from every other staircase edge,
but at distance at least $3m/2$ from any edge of a comb, a dihedral
rotation cannot cause a comb and the staircase to intersect.
Therefore, the only possible intersection during the rotation occurs
between the two combs.  Since the height of an edge is maintained
throughout the motion, intersections are only possible at the teeth.

Suppose we perform a dihedral rotation of angle $\pi$ at a vertical
staircase edge on the line $x=-b/2$.  This rotation reflects the right
comb across this vertical line, moving each tooth of the right comb
from $x$-coordinate $c'$ to $x$-coordinate $-c'-b$.  This rotation
causes two teeth to collide if and only if $a' = -c'-b$, or
equivalently, $a'+b+c'=0$, for some elements $a'\in A$ and $c'\in C$.

\begin{figure}[t]
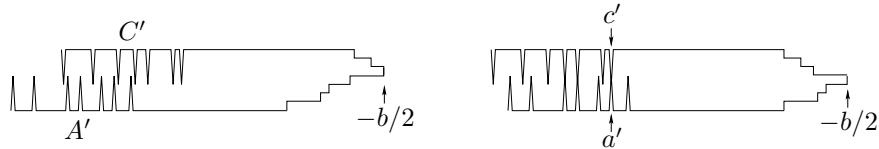

\centerline{\input threesumspins.pstex_t}
\caption{A dihedral rotation at $x=-b/2$.  \emph{Left:}~No collision
implies that $a'+b+c'\ne 0$ for all $a'\in A'$ and $c'\in C'$.
\emph{Right:}~A collision implies that $a' + b + c' = 0$ for some
$a'\in A'$ and $c'\in C'$.}
\label{fig-threesumspins}
\end{figure}

We perform a dihedral rotation query for each vertical staircase
edge.  If any of these rotations is infeasible, the infeasible
rotation identifies three elements $a'\in A'$, $b\in B$, and $c'\in
C'$ such that $a'+b+c' = 0$.  Conversely, if every dihedral rotations
are feasible, there is no such triple.  Thus, by performing at most
$n$ dihedral rotation queries, we solve the original instance of
\threesump.

Let $P(n)$ denote the time to preprocess a chain of $n$ segments for
static dihedral rotation queries, and let $Q(n)$ be the worst-case
time for a single query.  Our reduction solves any instance of
\threesump\ of size $n$ in time $O(n\log n) + P(n) + n Q(n)$.  Results
of Ben-Or~\cite{b-lbact-83} imply that $\threesum(n) = \Omega(n\log
n)$.  It follows that $P(n) + n Q(n) = \Omega(\threesum(n))$, as
desired.
\end{proof}



\section{Dynamic dihedral rotation queries}
\label{sec-mesq}

We now switch our attention to the case of \emph{dynamic} dihedral
rotation queries.  Given an edge $e$ and an angle $\phi$, a dynamic
dihedral rotation query determines whether the dihedral rotation at
edge~$e$ by angle $\phi$ is feasible, and if it is, modifies the chain
by performing the rotation.  These queries allow us to determine the
feasibility of an arbitrary sequence of rotations.  For example, we
might ask, ``Can we rotate at edge $e_1$ by angle $\phi_1$, then edge
$e_2$ by angle $\phi_2$, then edge $e_3$ by angle $\phi_3$, without
any collisions at any time?''

Dynamic dihedral rotation queries are more general than the static
queries considered earlier, since we can simulate any static query
using at most two dynamic queries.  Specifically, if a rotation at
edge $e$ by angle $\phi$ is feasible, a second rotation at edge $e$
with angle $-\phi$ restores the chain to its original configuration.
Thus, any lower bound for static queries automatically applies (up to
a constant factor) to dynamic queries as well.  However, we conjecture
that dynamic queries are much harder.

\begin{conjecture}
\label{conj:dynamic}
In any scheme to preprocess a chain of $n$ edges to answer dynamic
dihedral rotation queries, the worst-case query time is $\Omega(n)$,
regardless of the preprocessing time.
\end{conjecture}

One might reasonably ask why Conjecture \ref{conj:dynamic} is in any
way nontrivial; after all, a dihedral rotation can change the
locations of up to $n-1$ vertices of the chain.  However, there is no
reason \emph{a~priori} that we need to modify these locations
explicitly.  In fact, if we do not care about collisions, we can
perform any sequence of dihedral rotations, each in $O(\log n)$ time,
using a simple, linear-size data structure.

\begin{theorem}
Given a chain of $n$ edges and a sequence of $k$ dihedral rotations,
all assumed to be feasible, we can compute the resulting chain in $O(n
+ k\log n)$ time and $O(n)$ space.
\end{theorem}

\def\MM{\overline{M}}

\begin{proof}
We maintain a balanced binary tree $T$ whose leaves represent the
vertices of the chain and whose internal nodes represent contiguous
subchains.  At each leaf $\ell$, we store a set of
$(x,y,z)$-coordinates for the corresponding chain vertex $p_\ell$.  At
every node $v$, we store some representation for a rigid motion
$M_v:\Real^3\to\Real^3$.  The actual coordinates of each vertex are
computed by composing all the transformations stored on the
corresponding root-to-leaf path.  Specifically, for each tree node
$v$, we define the function $\MM_v:\Real^3\to\Real^3$ as follows.  If
$v$ is the root, then $\MM_v = M_v$; otherwise, $\MM_v = M_v \circ
\MM_u$, where $u$ is the parent of~$v$.  Finally, if leaf $\ell$
stores the coordinates $(x,y,z)$, then the actual location of the
corresponding chain vertex $p_\ell$ is $\MM_\ell(x,y,z)$.

Initially, every $M_v$ is the identity transformation, and each leaf
stores the actual coordinates of its chain vertex.  We can easily
create the initial tree in $O(n)$ time.

Now suppose we want to perform a dihedral rotation at some edge $e$ by
angle $\phi$.  Let $R(\phi,e)$ denote the rigid motion that rotates
space around the line through $e$ by angle $\phi$.  We want to apply
this transformation to the subchain on one side of the edge $e$.  To
do this, we first find a set of $O(\log n)$ maximal subtrees of $T$
containing the vertices of this subchain, in $O(\log n)$ time.  These
subtrees can be found using a binary search for one endpoint of $e$ in
$O(\log n)$ time.  Then, for each root $v$ of one of the maximal
subtrees, we replace $M_v$ with the composition $R(\phi,e) \circ M_v$;
this has the effect of replacing $\MM_v$ with $R(\phi,e) \circ \MM_v$.

Finally, after all $k$ rotations have been performed, we can recover
the actual coordinates of the chain vertices in $O(n)$ time by a
simple tree traversal.
\end{proof}

In support of Conjecture \ref{conj:dynamic}, we prove in this section
that sublinear dynamic dihedral rotation queries are impossible unless
there is a \emph{nonuniform} family of algorithms for \threesum\ with
subquadratic running time.  A nonuniform family of algorithms consists
of an infinite sequence of algorithms, one for each possible input
size, not necessarily described by a single efficient procedure.  The
existence of such a family seems unlikely in light of Erickson's
$\Omega(n^2)$ lower bound for \threesum, although in a restricted
model of computation~\cite{e-lblsp-98}.  Even if such a nonuniform
family of algorithms does exist, our preprocessing time would be at
least the time required to construct the $n$th algorithm in the
family.

The distinction between uniform and nonuniform algorithms is best
illustrated by a result of Meyer auf der Heide~\cite{m-ptlsa-84}, who
proved that for each input length~$n$, there is a linear decision tree
of depth $O(n^4 \log n)$ that solves the (NP-complete)
\textsc{Knapsack} problem: Given a set of $n$ real numbers, does any
subset sum to $1$?  These linear decision trees exploit `hardwired'
information that any uniform algorithm would require superpolynomial
time to compute on the fly, unless P=NP.\footnote{Specifically, the
computation path for any input implicitly depends on which subset of a
set of $2^n$ hyperplanes intersects a cell in a grid of hypercubes in
$\Real^n$.  Although we can locate the appropriate cell on the fly in
polynomial time, calculating the subset of hyperplanes that intersect
it is NP-hard.}  Of course, one could precompute all this hardwired
information if the input size $n$ is known in advance, but this would
require exponential time and space (even if P=NP).




We now demonstrate the link between a nonuniform algorithms for
\threesump\ and the dynamic dihedral rotation query problem.  Let
$\threesumprep(n)$ denote the time required to construct, given the
input size $n$, an algorithm that can solve any instance of
\threesump\ of size $n$ in $o(n^2)$ time.  For example, if there is a
nonuniform family of linear decision trees of subquadratic depth,
$\threesumprep(n)$ is (at most) the time to construct the $n$th tree
in the family.  If there is no subquadratic nonuniform algorithm for
\threesump, then $\threesumprep(n) = \infty$.

\begin{theorem}
\label{theo-lbmultipleedgespins}
Suppose we have a data structure that can answer dynamic dihedral
rotation queries in $Q(n)$ time, after $P(n)$ preprocessing time, for
any chain of length $n$.  Then either $Q(n) = \Omega(n)$, or $P(n) =
\Omega(\threesumprep(n))$, or $\threesum(n) = o(n^2)$.
\end{theorem}


\begin{proof}
We reduce the construction of a subquadratic nonuniform algorithm for
\threesump\ on sets of size $n$ to a series of dynamic dihedral
rotation queries as follows.  Suppose we are given the integer~$n$ and
asked to construct an algorithm for any \threesump\ problem where each
set has $n$ elements.  We create a chain whose structure is determined
solely by the number $n$, and spend time $P(n)$ preprocessing it to
answer dynamic dihedral rotation queries.  When the preprocessing has
finished, the sets $A$, $B$, and $C$ are revealed.  We then perform a
sequence of $O(n)$ dihedral rotations, each in time $Q(n)$, that move
the chain into a configuration similar to the one in
Figure~\ref{fig-threesum} for the three sets.  After $O(n)$ additional
rotations, as in the proof of Theorem~\ref{theo-lbedgespinquery}, the
given instance of \threesump\ is solved.  If $Q(n) = o(n)$, then the
\threesump\ instance has been solved in subquadratic time.  Thus,
constructing a subquadratic algorithm for instances of \threesump\ of
size $n$ has been reduced to constructing and preprocessing the chain.
It follows that $P(n)$ must be $\Omega(\threesumprep(n))$.  In
particular, if there is no subquadratic nonuniform algorithm for
\threesump, then $Q(n)$ must be $\Omega(n)$.

For any positive integer $n$, we construct a \emph{canonical} planar
chain as follows.  We begin by building a chain consisting of a left
comb pointing up, a staircase, and a right comb pointing down, exactly
as in the previous section.  See Figure~\ref{fig-potschain}.  Each
comb consists of $n$ teeth, each of height $1$, where adjacent pairs
of teeth are distance $2$ apart.  The staircase consists of $n+1$
steps, each with width $1$ and height $2/n$.  The distance between the
staircase and either comb is $7$.

\begin{figure}
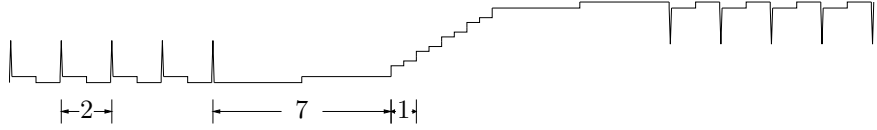

\centerline{\input potschain-ltx.pstex_t}
\caption{The canonical chain for $n=5$; see the proof of
Theorem~\ref{theo-lbmultipleedgespins}.} 
\label{fig-potschain}
\end{figure}

We then replace every horizontal segment in the chain with a
\emph{hinge} consisting of five segments, as shown in
Figure~\ref{fig-threesumgadget}.  Each hinge allows us to bring any
adjacent pair of vertical segments arbitrarily close together by a
short sequence of dihedral rotations.  Specifically, referring to the
left side of Figure~\ref{fig-threesumgadget}, we can bring teeth
$a'_1$ and $a'_2$ to any desired distance by performing a dihedral
rotation at $\alpha_1$ by some angle $0<\theta<\pi/2$, a dihedral
rotation at $uv$ by $-2\theta$, and a dihedral rotation at $\alpha_2$
by angle~$\theta$.  After the three rotations, the teeth are at any
desired distance less than $1$, and the rest of the chain is
unaffected except for a translation.  We easily verify that if the
portion of the chain on one side of the hinge is coplanar, then we can
perform these rotations without collisions.

\begin{figure}
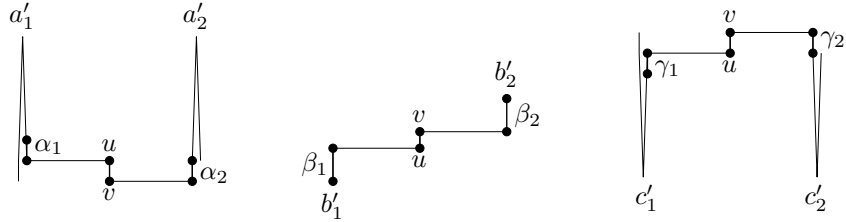

\centerline{\input threesumgadget-ltx.pstex_t}
\caption{Hinges for the left comb, the staircase, and the right comb.}
\label{fig-threesumgadget}
\end{figure}

Once we construct the canonical planar chain, we preprocess it for
dynamic dihedral rotation queries in time $P(n)$.

Now suppose we are given three sets $A$, $B$, and $C$, each containing
$n$ integers, and are asked if they respectively contain three
elements $a$, $b$, and $c$ whose sum is zero.  To solve this instance
of \threesump, we perform a sequence of $O(n)$ dynamic dihedral
rotation queries; a triple of elements summing to zero exists if and
only if some dihedral rotation in this sequence is infeasible.

Our reduction will be easier if we assume that the three input sets
$A$, $B$, and $C$ have the same number of elements.  If some set has
fewer elements than another, then we can augment the smaller set with
elements of the form $i/4n$, for some small integer $i$, without
affecting the outcome of \threesump.  (Because the other elements of
the sets are integers, none of these fractions can contribute to a
triple of elements that sum to zero).  We will also assume, as in the
proof of Theorem~\ref{theo-lbedgespinquery}, that the sets are given
in sorted order.

Let $m$ be the maximum absolute value of any element in $A \cup B \cup
C$.  We define three new sets $A'$, $B'$, and $C'$ as follows:
\[
	A' = \set{a/m - 5 \mid a\in A}, \qquad
	B' = \set{b/2m \mid b\in B}, \qquad
	C' = \set{c/m + 5 \mid c\in C}.
\]
Clearly, the original sets contain elements $a,b,c$ such that
$a+b+c=0$ if and only if these new sets contain corresponding elements
$a',b',c'$ such that $a'+c'=2b'$.

To encode these sets into our chain, we manipulate the hinges in order
from left to right so that the $x$-coordinates of the left comb's
teeth are the elements of $A'$, the $x$-coordinates of the vertical
staircase edges are the elements of $B'$, the $x$-coordinates of the
right comb's teeth are the elements of~$C'$.  This manipulation is
always possible, because the required distance between the ends of any
hinge is no more than their distance in the original canonical chain.
An example of the final configuration is in
Figure~\ref{fig-potsfinal}.

\begin{figure}
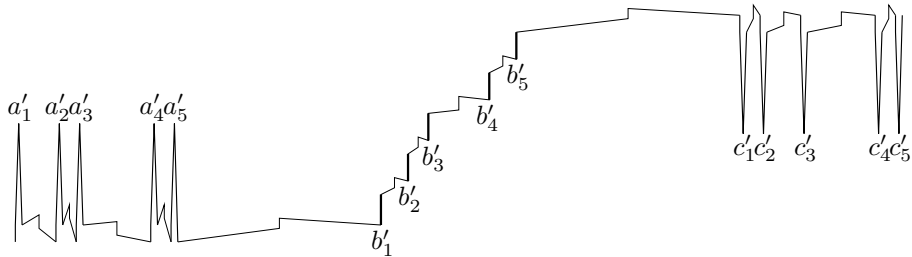

\centerline{\input potschainset2-ltx.pstex_t}
\caption{The canonical chain, manipulated to encode $A'$, $B'$, and $C'$.
\emph{(Shown in stereo in Figure~\ref{fig-stereo-potsfinal}).}}
\label{fig-potsfinal}
\end{figure}

\begin{figure}
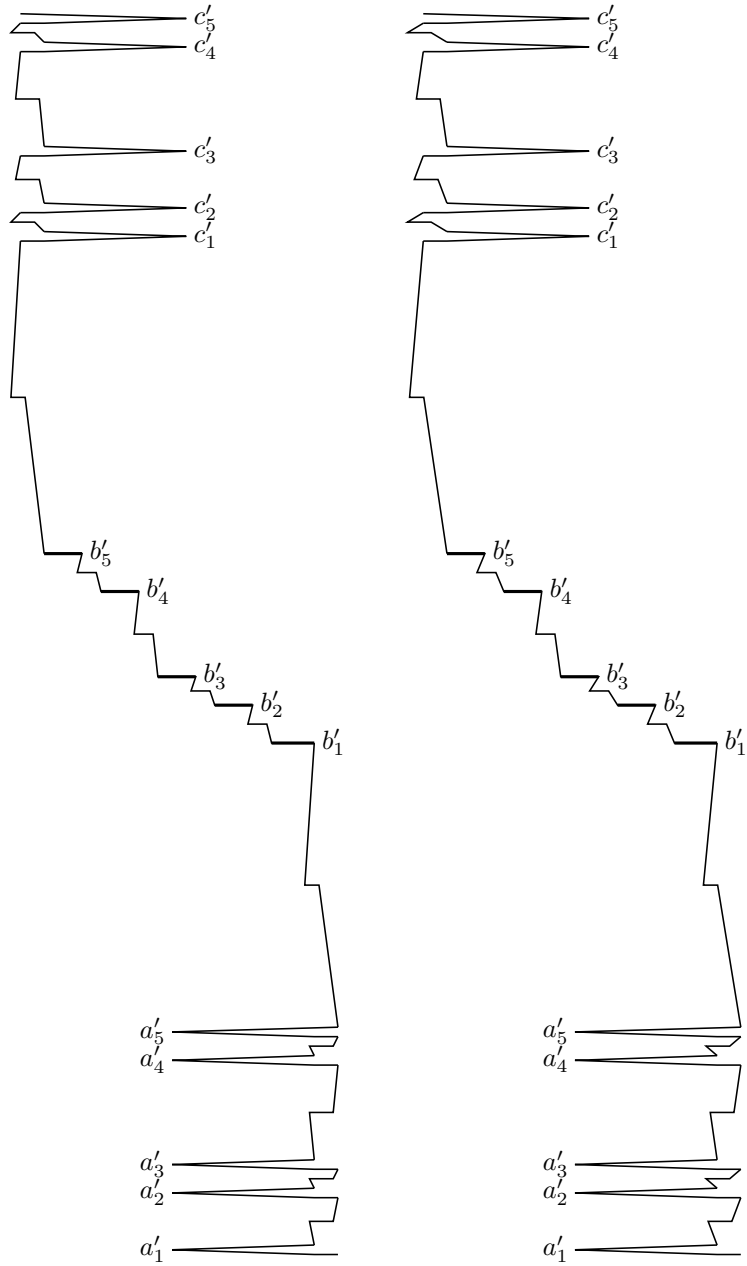

\centerline{\input stereo-potschainset-side-ltx.pstex_t}
\caption{In stereo:  The canonical chain, manipulated to encode $A'$,
$B'$, and $C'$.}
\label{fig-stereo-potsfinal}
\end{figure}

We observe that the chain does not self-intersect during these
dihedral rotations by examining the hinges in
Figure~\ref{fig-threesumgadget}.  As described earlier, each hinge is
manipulated using a sequence of three rotations.  Because we
manipulate the hinges in order from left to right, whenever we flex a
hinge, the portion of the chain to the right of that hinge is
coplanar.

Once the chain is set for $A'$, $B'$, and $C'$, we perform dihedral
rotations of angle $2\pi$ at every vertical edge in the staircase that
corresponds to an element of $B'$.  Just as in the proof of
Theorem~\ref{theo-lbedgespinquery}, the chain self-intersects if and
only if there exists a triplet $a' - 2b' + c' = 0$.  Thus, the
sequence of $n$ dynamic queries solves the original \threesump\
problem.

We spent time $P(n)$ preprocessing the chain before the sets were
revealed, and time $10n Q(n)$ for $10n$ dihedral rotations---$9n$
rotations to set the $3n$ hinges, and $n$ more to test for collisions.
Thus, after $P(n)$ preprocessing time, we can answer any instance of
\threesump\ of size $n$ in time $O(n Q(n))$.  The theorem now follows
immediately.
\end{proof}

\subsection*{Acknowledgments}

We would like to thank Godfried Toussaint for organizing the Workshop
on Computational Polygonal Entanglement Theory on February 4-11, 2000,
at which this research was initiated.  We also thank the other
participants of the workshop, Oswin Aichholzer, David Bremner, Carmen
Cort\'es, Erik Demaine, Vida Dujmovi\'c, Ferran Hurtado, Henk Meijer,
Bel\'en Palop, Suneeta Ramaswami, Vera Sacrist\'an, and Godfried
Toussaint for stimulating conversations.

\bibliographystyle{plain}
\bibliography{geom,linkrefs,../../custom/polymerrefs}
\end{document}